\documentstyle[12pt,epsf]{article}
\if@twoside                                     
\else
\fi
\title{Symmetry, Local Linearization, and Gauge Classification of the
  Doebner-Goldin Equation}
\author{P.~Nattermann \\
  \small Institute for Theoretical Physics A\\[-1ex]
  \small Technical University Clausthal\\[-1ex]
  \small D-38678 Clausthal-Zellerfeld, Germany\\[-1ex]
  \small E-mail: {\tt aspn@pta3.pt.tu-clausthal.de}}
\date{}
\font\teneufm=eufm10                            
\font\seveneufm=eufm7
\font\fiveeufm=eufm5

\newfam\eufmfam
\textfont\eufmfam=\teneufm
\scriptfont\eufmfam=\seveneufm
\scriptscriptfont\eufmfam=\fiveeufm

\let\goth\frak

\def\ds{\displaystyle}
\def\sss{\scriptscriptstyle}
\let\epsilon\varepsilon
\def\rz{\ifmmode{I\hskip -3pt R}
    \else{\hbox{$I\hskip -3pt R$}}\fi}
\def\cz{\ifmmode{C\hskip-4.8pt\vrule height5.8pt\hskip6.3pt}
    \else{\hbox{$C\hskip-4.8pt\vrule height5.8pt\hskip6.3pt$}}\fi}
\def\gz{\ifmmode{Z\hskip -4.8pt Z}
    \else{\hbox{$Z\hskip -4.8pt Z$}}\fi}
\def\sfrac#1#2{\mbox{$\frac{#1}{#2}$}}
\def\splus{{\subset \!\!\!\!\!\! +}}
\def\stimes{{\subset \!\!\!\!\!\! \times}}
\def\Dt{\frac{d}{dt}}
\def\dt{\partial_t}
\def\x{\vec{x}}
\def\F{\goth{F}}
\def\Diff{\mbox{\it Diff}}
\def\Vect{\goth{X}}
\def\grad{\vec{\nabla}}
\def\Tr{\mbox{Tr}}
\def\op#1{\mbox{\boldmath $#1$}}
\def\Ref#1{(\ref{#1})}
\def\Aff{\mbox{\it Aff\/}}
\def\aff{\mbox{\it aff\/}}
\def\pr2{\mbox{pr}^{\sss (2)}\!}
\def\half{\sfrac{1}{2}}
\def\bbbone{{\mathchoice {1\mskip-4mu \mbox{l}} {1\mskip-4mu \mbox{l}}
{1\mskip-4.5mu \mbox{l}} {1\mskip-5mu \mbox{l}}}}

\begin{document}
{\let\newpage\relax
\hfill {\bf ASI-TPA/8/95}\\
\mbox{}\hfill {\sl to appear in Reports on Mathematical Physics}
\maketitle}
\begin{abstract}
\noindent
For the family of nonlinear {\sc Schr\"odinger} equations
derived by {\sc H.-D.~Doebner} and {\sc G.A.~Goldin}
(J.~Phys.~A {\bf 27}, 1771) we calculate the complete set
of {\sc Lie} symmetries. For various subfamilies we find different
finite and infinite dimensional {\sc Lie} symmetry algebras. Two of
the latter lead to a local transformation
linearizing the particular subfamily. One type of these transformations
leaves the whole family of equations invariant, giving
rise to a gauge classification of the family.
The {\sc Lie} symmetry algebras and their corresponding subalgebras
are finally characterized by gauge invariant parameters.
\end{abstract}

\section{Introduction}
In a series of articles \cite{DoeGol1}\nocite{DoeGol2,DoeGol3}--\cite{DoeGol4}
{\sc H.-D.~Doebner} and {\sc G.A.~Goldin} derived a family of nonlinear
{Schr\"odinger} equations on $\rz^n$ from the representation
theory of the semi-direct product of the group $\Diff(\rz^n)$
of diffeomorphisms on
$\rz^n$ and the {\sc Abel}ian group $C^\infty(\rz^n)$
of smooth functions on $\rz^n$,
viewed as the \lq kinematical symmetry group\rq\ of the configuration
space $\rz^n$.
Their basic observation is that unitarily inequivalent representations
of its {\sc Lie} algebra ${\cal S}(\rz^n)=\Vect(\rz^n)\splus
C^\infty(\rz^n)$ are labeled by a real number
$c$, which they interpreted as a new quantum number. On
${\cal H}=L^2(\rz^n,d^n\!x)$ the representatives $\op{Q}(f)$ of functions $f\in
C^\infty(\rz^n)$ and $\op{P}^c(\vec{g})$ of vector-fields $\vec{g}\in
\Vect(\rz^n)$ are
\begin{equation}
\begin{array}{rcl}
  \ds \left(\op{Q}(f)\psi\right) (\vec{x}) &=& \ds
    f(\vec{x})\psi(\vec{x})\,,  \\
  \ds \left(\op{P}^c(\vec{g})\psi\right) (\vec{x}) &=& \ds
    \frac{\hbar}{i}\vec{g}(\vec{x})\cdot
    \left(\grad\psi\right)(\vec{x}) + \left(\frac{\hbar}{2i}+c\right)
    \left(\grad\!\cdot\!\vec{g}\right)(\vec{x})\, \psi(\vec{x})\,.
\end{array}
\end{equation}
A connection between these representations of the kinematical
algebra ${\cal S}(\rz^n)$ and a consistent time evolution of the quantum state
can be established as follows \cite{Natter1,DoeHen1}\footnote{The
representations and the derivation can be generalized to smooth
configuration manifolds.}:
The representation of ${\cal S}(\rz^n)$ by functions on the classical phase
space $\rz^n\times\rz^n$,
\begin{equation}
\begin{array}{rcl}
  \ds Q_f (\vec{x},\vec{p}) &=& \ds
    f(\vec{x})\,,  \\
  \ds P_{\vec{g}}(\vec{x},\vec{p}) &=& \ds
    \vec{p}\cdot \vec{g}(\vec{x})\,,
\end{array}
\end{equation}
for a classical system with symmetric and invertible mass matrix $M$,
fulfills the dynamical relation
\begin{equation}
  \Dt Q_f(\vec{x}(t),\vec{p}(t)) = P_{M^{-1}\grad
    f}(\vec{x}(t),\vec{p}(t)) \,.
\end{equation}
In analogy to this equations a natural requirement for the time
evolution of a (mixed) quantum state $\op{W}(t)\in{\cal T}_+^1({\cal
  H})$ is a kind of generalized first {\sc Ehrenfest} relation
\begin{equation}
  \Dt \Tr\left(\op{W}(t)\op{Q}(f)\right) = \Tr\left(\op{W}(t)
    \op{P}^c ( M^{-1} \grad f)\right) \,.
\label{qmdyn:eq}
\end{equation}
While there are evolution equations of {\sc Lindblad} type
\cite{Lindbl1} for mixed states fulfilling this equation, we restrict
ourselves here to time evolutions of pure states, i.e.\ we impose the
additional constraint $\op{W}(t) = |\psi(t)\rangle\langle\psi(t)|$. For a
single particle with mass $m$, $M=m\bbbone$, equation \Ref{qmdyn:eq}
leads to an evolution equation of {\sc Fokker-Planck} type for the
quantum mechanical probability and current densities
$\rho=\psi\bar\psi$ and $\vec{j}=\frac{\hbar}{2im}
(\bar\psi\grad\psi-\psi\grad\bar\psi)$:
\begin{equation}
  \dt \rho = -\grad\cdot \vec{j} + D \Delta \rho,
\end{equation}
where $D:= \frac{c}{m}$ is a diffusion parameter.
The evolution equation of the wave function $\psi\in L^2(\rz^n,d^n\!x)$
is restricted to a nonlinear Schr\"odinger equation
\begin{equation}
  i\hbar\dt \psi = \op{H}\psi
    +i\half \hbar D \frac{\Delta\rho}{\rho}\psi +R[\psi]\psi\,,
\end{equation}
where $\op{H}=-\frac{\hbar^2}{2m}\Delta+V$ is a
{\sc Hamilton} operator and $R[\psi]$ an arbitrary real-valued functional
of $\psi$.
Requiring for the real functional $R[\psi]$ some of the properties of the
imaginary functional $i\half \hbar D \frac{\Delta\rho}{\rho}$
finally leads to a five parameter family
of nonlinear {\sc Schr\"odinger} equations, the general {\sc Doebner-Goldin}
equation in $n+1$ dimensions:
\begin{equation}
  i\hbar\dt \psi = \op{H}\psi
    +i\half \hbar D \frac{\Delta\rho}{\rho}\psi
    +\hbar D\sum_{j=1}^5 c_j R_j[\psi]\psi\,.
\label{GDGE:eq}
\end{equation}
Here the nonlinear functionals $R_j[\psi]$ are complex homogeneous of
degree zero,
\begin{equation}
\begin{array}{c}
\ds  R_1[\psi] := \frac{m}{\hbar}\frac{\grad\cdot\vec{j}}{\rho}\,,\qquad
  R_2[\psi] := \frac{\Delta\rho}{\rho}\,,\qquad
  R_3[\psi] := \frac{m^2}{\hbar^2} \frac{\vec{j}^{\,2}}{\rho^2}\,, \\[1mm]
\ds  R_4[\psi] :=
  \frac{m}{\hbar}\frac{\vec{j}\cdot\grad\rho}{\rho^2}\,,\qquad
  R_5[\psi] := \frac{(\grad\rho)^2}{\rho^2}\,,
\end{array}
\end{equation}
and $c_j\in\rz$ are dimensionless \lq model\rq\ parameters. For the purpose
of this paper we will rewrite equation \Ref{GDGE:eq} completely in terms of
the real functionals $R_j[\psi]$, using the decomposition of the {\sc
  Laplace} operator
$\Delta\psi =(iR_1[\psi]+(1/2)R_2[\psi] -R_3[\psi]-(1/4)R_5[\psi])\,\psi$,
\begin{equation}
i\dt\psi = i\sum_{j=1}^2 \nu_j R_j[\psi]\psi
           +\sum_{j=1}^5 \mu_j R_j[\psi]\psi
+ \mu_0 V\psi \,,\qquad \nu_1\neq 0\,.
\label{nse:eq}
\end{equation}
Homogeneous nonlinear {\sc Schr\"odinger} equations of this type,
though with a different
choice of non-vanishing complex parameters $\mu_j+i\nu_j$, were introduced in
\cite{Sabati2}. In our case the free equation ($V\equiv 0$) can be viewed as a
seven parameter family of nonlinear Schr\"odinger
equations.

Several properties of the DG--equation have already been observed such as
its homogeneity and separability, the linearization and
{\sc Hamilton}ian formulation of a subfamily,
and stationary as well as some non-stationary solutions have been found
[2--5,8--15].
Nevertheless, apart from the invariance of a subfamily under {\sc
Galilei} transformations, the {\sc Lie} symmetry  of this partial
differential equation (PDE) was only investigated for small
nonlinearities of the $1+1$ dimensional DG--equation \cite{Natter3}.

In the case of the free linear {\sc Schr\"odinger}
equation (SE) the {\sc Lie} symmetry group is well known
and amounts to a projective unitary representation of the \lq {\sc
  Schr\"odinger} group\rq\ \cite{Nieder1,Hagen1,BarXu1}
\begin{equation}
  Sch(n) := \left( SL(2,\rz)\otimes SO(n) \right) \stimes
  \left(T(n)\otimes T(n)\right)\,,
\label{SchG:eq}
\end{equation}
where the action of $SO(n)$ on the {\sc Abel}ian groups $T(n)$
(representing
space translations and special {\sc Galilei} transformations) is as
usual and the elements of $SL(2,\rz)$
act on $T(n)\otimes T(n)$ as $2\!\times\!2$ matrices. Due to the linearity
of the SE there are two additional types of symmetry transformations:
on the one hand the multiplication of the wavefunction by a real
number (real homogeneity)\footnote{Note that the multiplication with a
phase factor is already contained in the projective representation of
$Sch(n)$.}, on the other hand the addition of an arbitrary solution of
the equation, which is in fact an infinite dimensional symmetry group.

In section~\ref{lie:sec} we determine the maximal {\sc Lie} symmetry
algebra of the {\em free\/} ($V\equiv 0$) DG--equation.
Exploiting the condition for a generator of a {\sc Lie}
symmetry in section~\ref{gen:sec} we are led to distinguish various
subfamilies. Among the {\sc Lie} symmetries there are {\em
finite\/} and an {\em infinite\/} dimensional ones and they will be
discussed separately in
sections~\ref{fin:sec} and~\ref{inf:sec}. It turns out that part of the
infinite dimensional symmetry corresponds to a linearization of a
subfamily of the DG--equation \cite{Natter2,AubSab1}. As the symmetry
algebra is maximal the
linearizing transformations obtained here are the only {\em local\/}
ones, i.e.\ the only ones depending purely on the values of the wave
functions.

One type of these transformations also leaves the
probability density invariant and has therefore been called
{\em nonlinear gauge transformation\/} \cite{DoGoNa1}.
As we will see in section~\ref{cla:sec} the whole family of
DG--equations is
invariant under these gauge transformations. Summarizing the results
we are finally led to a gauge invariant classification of the family
\Ref{GDGE:eq} and its symmetries.

\section{Lie symmetries}\label{lie:sec}
\subsection{Determining equations}\label{gen:sec}
The methods of finding {\sc Lie} symmetry algebras of PDEs have been
discussed in detail in the monographs
\cite{Ovsian1:book,Olver1:book,FuShSe1:book} and computer algebra
programs are available for the calculations, e.g.~\cite{Bauman1}. Here
we sketch roughly the procedure to fix our notation.

Using the polar decomposition for the wavefunction
$\psi(\x,t) = \exp[r(\x,t)+is(\x,t)]$, the free ($V\equiv 0$) complex
evolution equation \Ref{nse:eq} leads to two real evolution equations
for the real functions $r$ and $s$:
\begin{equation}
\begin{array}{rcl}
\ds \F_1(r_t,\grad r,\Delta r,s_t, \grad s, \Delta s)
  &:=& \ds
  -r_t +2\nu_2\Delta r +\nu_1 \Delta s
   + 4\nu_2 (\grad r)^2 + 2\nu_1 \grad r\cdot \grad s\\
  &=& 0\,, \\[1mm]
\ds  \F_2(r_t,\grad r,\Delta r,s_t, \grad s, \Delta s)
  &:=& \ds
  s_t + 2\mu_2 \Delta r + \mu_1 \Delta s
  + 4(\mu_2+\mu_5)(\grad r)^2 \\&& \ds
  +2(\mu_1+\mu_4)\grad r\cdot \grad s
  +\mu_3 (\grad s)^2 \\
  &=& 0\,,
\end{array}
\label{GDGEsys:eq}
\end{equation}
where we have used the usual indication of partial derivatives by subscripts.

A sufficient condition for the vector-field
\begin{equation}
  X = \sum_{j=1}^n\xi^{\sss (j)}(\x,t,r,s) \partial_{x_{\! j}} +
  \tau(\x,t,r,s) \partial_t
      +\phi(\x,t,r,s) \partial_r + \sigma(\x,t,r,s) \partial_s
\label{Xsym:eq}
\end{equation}
on the space of independent ($\x,t$) and dependent variables
($r,s$) to be a generator of a
symmetry group of transformations of the system \Ref{GDGEsys:eq}
described by
the map $\F=(\F_1,\F_2)$ is the vanishing action of its
second prolongation $\pr2 X$ on $\F$ restricted to
its solution variety $\Sigma:=\F^{-1}(0)$,
\begin{equation}
  \left(\pr2 X\, \F_j\right)_{|_\Sigma} = 0,\quad j=1,2\,.
\label{Xcond:eq}
\end{equation}
For the system \Ref{GDGEsys:eq} this condition is also
{\em necessary\/} since $\F$ is analytic and of
rank 2 on $\Sigma$ \cite[Theorem 2.71, Lemma 2.74]{Olver1:book}. Thus,
the calculation of all vector-fields $X$ fulfilling \Ref{Xcond:eq}
yields its maximal {\sc Lie} symmetry algebra.

Condition \Ref{Xcond:eq} leads to a set of linear PDEs called
{\em determining equations\/}
for the coefficients $\xi^{\sss (j)}$, $\tau$, $\phi$, and $\sigma$
of the vector-field $X$.
Since $\F$ depends only on the dependent variables $(r,s)$, their first
order derivatives, and
their Laplacian, calculations are reduced considerably.
A set of trivial equations among the determining equations yield
$\xi^{\sss (j)}=\xi^{\sss (j)}(\x,t)$ and $\tau=\tau(t)$.
With this restricted dependence the remaining
independent determining equations are
\begin{eqnarray}
\begin{array}[b]{r}
-\xi^{\sss (j)}_t
  - \mu_1 \Delta \xi^{\sss (j)}
  + 2 (\mu_1+\mu_4) \phi_{x_{\! j}}
  + 4 \mu_2  \phi_{{x_{\! j}}s}
  + 2 \mu_3 \sigma_{x_{\! j}}
  + 2 \mu_1  \sigma_{{x_{\! j}}s}
\end{array}
&=& 0,\label{d6:eq} \\[1mm]
- \nu_1 \Delta \xi^{\sss (j)}
  + 2 \nu_1 \phi_{x_{\! j}}
  + 4 \nu_2  \phi_{{x_{\! j}}s}
  + 2 \nu_1 \sigma_{{x_{\! j}}s}
&=& 0, \label{d7:eq} \\[1mm]
\begin{array}[b]{r}
  - \mu_2 \Delta \xi^{\sss (j)}
  + 4 (\mu_2 + \mu_5) \phi_{x_{\! j}}
  + 2 \mu_2 \phi_{{x_{\! j}}r}
  + (\mu_1 + \mu_4) \sigma_{x_{\! j}}
  + \mu_1 \sigma_{{x_{\! j}}r}
\end{array}
&=& 0, \label{d8:eq} \\[1mm]
\xi^{\sss (j)}_t
- 2 \nu_2 \Delta \xi^{\sss (j)}
+ 8 \nu_2 \phi_{x_{\! j}}
+ 4 \nu_2 \phi_{{x_{\! j}}r}
+ 2 \nu_1 \sigma_{x_{\! j}}
+ 2 \nu_1 \sigma_{{x_{\! j}}r}
&=& 0, \label{d9:eq} \\[1mm]
\sigma_t
  + \mu_1 \Delta\sigma
  + 2 \mu_2 \Delta\phi
&=& 0, \label{d10:eq} \\[1mm]
-\phi_t
  + 2 \nu_2 \Delta\phi
  + \nu_1 \Delta\sigma
&=& 0, \label{d11:eq} \\[1mm]
\xi^{\sss (j)}_{x_k} + \xi^{\sss (k)}_{x_{\! j}}
&=& 0, \label{drot:eq} \\[1mm]
\begin{array}[b]{r}
2 (\mu_1 + \mu_4) \phi_s
  + 2 \mu_2 \phi_{ss}
  + \mu_3 \sigma_s
  + \mu_1 \sigma_{ss}
  + \mu_3 \tau_t
  - 2 \mu_3 \xi^{\sss (j)}_{x_{\! j}}
\end{array}
&=& 0, \label{d12:eq} \\[1mm]
2 \mu_2 \phi_s
  + \nu_1 \sigma_r
  + \mu_1 \tau_t
  - 2 \mu_1 \xi^{\sss (j)}_{x_{\! j}}
&=& 0, \label{d13:eq} \\[1mm]
(\mu_1 + 2 \nu_2) \phi_s
  - \nu_1 \phi_r
  + \nu_1 \sigma_s
  + \nu_1 \tau_t
  - 2 \nu_1 \xi^{\sss (j)}_{x_{\! j}}
&=& 0, \label{d14:eq} \\[1mm]
\begin{array}[b]{r}
 4 (\mu_2+ \mu_5) \phi_s
  + (\mu_1+\mu_4) \phi_r
  + 2 \mu_2 \phi_{rs}\\
  + (\mu_3+\nu_1) \sigma_r
  + \mu_1 \sigma_{rs}
  + (\mu_1+\mu_4) \tau_t
  - 2 (\mu_1+\mu_4) \xi^{\sss (j)}_{x_{\! j}}
\end{array}
&=& 0, \label{d15:eq} \\[1mm]
\begin{array}[b]{r}
2 \mu_2 \phi_r
  - 2 \mu_2 \sigma_s
  + (\mu_1 + 2 \nu_2) \sigma_r
  + 2 \mu_2 \tau_t
  - 4 \mu_2 \xi^{\sss (j)}_{x_{\! j}}
\end{array}
&=& 0, \label{d16:eq} \\[1mm]
2 \mu_2 \phi_s
  + \nu_1 \sigma_r
  + 2 \nu_2 \tau_t
  - 4 \nu_2 \xi^{\sss (j)}_{x_{\! j}}
&=& 0, \label{d17:eq} \\[1mm]
 (\mu_1 + \mu_4 + 4 \nu_2) \phi_s
  + 2 \nu_2 \phi_{rs}
  + \nu_1 \sigma_s
  + \nu_1 \sigma_{rs}
  + \nu_1 \tau_t
  - 2 \nu_1 \xi^{\sss (j)}_{x_{\! j}}
&=& 0, \label{d18:eq} \\[1mm]
\begin{array}[b]{r}
8 (\mu_2+\mu_5) \phi_r
  - 4 (\mu_2+\mu_5) \sigma_s\\
  + 2 (\mu_1 +\mu_4+2 \nu_2) \sigma_r
  + \mu_1 \sigma_{rr}
  + 4 (\mu_2+\mu_5) \tau_t
  - 8 (\mu_2+\mu_5) \xi^{\sss (j)}_{x_{\! j}}
\end{array}
&=& 0, \label{d19:eq} \\[1mm]
\begin{array}[b]{r}
4 (\mu_2 + \mu_5) \phi_s
  + 4 \nu_2 \phi_r
  + 2 \nu_2 \phi_{rr}
  + 2 \nu_1 \sigma_r
  + \nu_1 \sigma_{rr}
  + 4 \nu_2 \tau_t
  - 8 \nu_2 \xi^{\sss (j)}_{x_{\! j}}
\end{array}
&=& 0, \label{d20:eq} \\[1mm]
\mu_3 \phi_s
  + 2 \nu_1 \phi_s
  + 2 \nu_2 \phi_{ss}
  + \nu_1 \sigma_{ss}
&=& 0. \label{d21:eq}
\end{eqnarray}
Combining the derivative of equation \Ref{d12:eq} with
respect to $s$ with equation \Ref{d21:eq} we obtain a linear second
order differential equation for $\phi_s$ as a function of $s$,
\begin{equation}
  \alpha_1\phi_{sss} + \alpha_2 \phi_{ss} + \alpha_3 \phi_{s} = 0,
\label{dphis:eq}
\end{equation}
with constant coefficients
$\alpha_1=2(\nu_1\mu_2-\nu_2\mu_1)$,
$\alpha_2= 2\nu_1\mu_4-\mu_3(\mu_1+2\nu_2)$, and
$\alpha_3= -\mu_3(\mu_3+2\nu_1)$.
Another ordinary differential equation of this type is obtained from
equations \Ref{d14:eq}, \Ref{d17:eq}, \Ref{d21:eq}, and their
differential consequences:
\begin{equation}
  \beta_1 \phi_{ss} + \beta_2 \phi_{s} = 0,
\label{dphis1:eq}
\end{equation}
where $\beta_1 = \nu_1(\mu_4+2\nu_2)-\mu_1(\mu_3+\nu_1)$ and $\beta_2
= (\mu_3+2\nu_1)(\mu_3+\nu_1)$.

Thus, for the general $s$--dependence of $\phi$ we have to distinguish
four different cases. First we have a polynomial dependence of $\phi$
on $s$ of maximal order two, if at most one of the $\alpha_j$ and one
of the
$\beta_j$ is non-vanishing. Then there are exponential solutions of
\Ref{dphis:eq} and \Ref{dphis1:eq} (including oscillating solutions),
as well as aperiodic solutions of \Ref{dphis:eq}. In all these cases
the $s$--dependence of $\sigma$ is obtained by an integration of
\Ref{d21:eq}. Finally, if all coefficients vanish,
$\alpha_j=\beta_k=0$, there is no information on the $s$--dependence
of $\phi$ and $\sigma$, but instead, using equation \Ref{d15:eq} as
well,
we get a restriction on the parameters:
\begin{equation}
  \mu_2 = \frac{\nu_2 \mu_1}{\nu_1}\,,\quad
  \mu_3 = -2\nu_1\,,\quad
  \mu_4 = -2\nu_2-\mu_1\,,\quad
  \mu_5 = -\frac{\nu_2 \mu_1}{\nu_1}\,.
\label{InfSub:eq}
\end{equation}

Since the solutions and their derivatives in the different cases are
functionally independent and the determining equations are linear we
can investigate them independently. Calculations for the different
cases show that in general we have two kinds of symmetries, a set of
{\em finite\/} dimensional {\sc Lie} symmetries arising from
polynomial or exponential solutions, and a set of {\em infinite\/}
dimensional {\sc Lie} symmetries connected with exponential
solutions or the subfamily \Ref{InfSub:eq}. The aperiodic
solution turns out to be consistent only for this subfamily as well.

We will discuss the finite and infinite symmetries separately in the
next two sections.

\subsection{Finite symmetries}\label{fin:sec}
Let us begin with the polynomial solutions of \Ref{dphis:eq} and
\Ref{dphis1:eq}, i.e.
\begin{equation}
  \phi(\x,t,r,s) = \phi^{\sss (2)}(\x,t,r) s^2
    + \phi^{\sss (1)}(\x,t,r)s +\phi^{\sss (0)}(\x,t,r)\,,
\label{phi_poly:eq}
\end{equation}
in which case $\sigma$ integrates to
\begin{equation}
  \sigma(\x,t,r,s) =
  \begin{array}[t]{l}
  -\frac{\mu_3+2\nu_1}{3\nu_1}
    \phi^{\sss (2)}(\x,t,r) s^3
  -\left(\frac{\mu_3+2\nu_1}{2\nu_1} \phi^{\sss (1)}(\x,t,r)
    +\frac{2\nu_2}{\nu_1} \phi^{\sss (2)}(\x,t,r) \right) s^2\\
  +\left( \sigma^{\sss (1)}(\x,t,r) - \frac{\mu_3+2\nu_1}{\nu_1}
    \phi^{\sss (0)}(\x,t,r) -\frac{2\nu_2}{\nu_1} \phi^{\sss
      (1)}(\x,t,r)\right) s + \sigma^{\sss (0)}(\x,t,r)
  \end{array}
\label{sigma_poly:eq}
\end{equation}
Neglecting the special case \Ref{InfSub:eq}, which will be treated in
the next section, we obtain from equations \Ref{d13:eq}--\Ref{d20:eq}
\begin{equation}
  \phi^{\sss (1)}\equiv \phi^{\sss (2)} \equiv
  \phi^{\sss (0)}_r \equiv \sigma^{\sss (1)}_r \equiv 0
  \quad\mbox{ and }\quad
  \sigma^{\sss (0)}_r \equiv -\frac{2\nu_2}{\nu_1}\left( \tau_t -2
    \xi^{(j)}_{x_{\!j}} \right)\,.
\end{equation}
With these results equations \Ref{d6:eq}--\Ref{drot:eq} lead to a
polynomial dependence of the coefficients of the vector-field $X$ on
all variables:
\begin{equation}
\begin{array}{rcl}
  \vec{\xi}(\x,t) &=& k_1\,t\x + k_2\, \x + A \x
  +\vec{b}\,t +\vec{a} \,, \\[1mm]
  \tau(t) &=& k_1 t^2 + (2k_2-k_6) t + k_3\,,\\[1mm]
  \phi(t,f)&=& -\frac{n}{2} k_1\, t + k_4 \,,\\[1mm]
  \sigma(\x,t) &=& \frac{2\nu_2}{\nu_1}k_6\,r + k_6\,s
    -\frac{1}{4\nu_1}k_1\, \x^2
    -\frac{1}{2\nu_1} \vec{b}\cdot\x
    +\frac{n\mu_1}{2\nu_1} k_1\, t + k_5 -\frac{2\nu_2}{\nu_1} k_4 \,,
\end{array}
\end{equation}
where $A$ is a real antisymmetric matrix and $k_j,\vec{a},\vec{b}$
are real constants with the additional conditions
\begin{eqnarray}
  k_1=0\,, \vec{b}=\vec{0} &\mbox{ or }& \mu_1+\mu_4 =
  \mu_3+\nu_1=0\,,
\label{GalSub:eq}\\
  k_6 =0 &\mbox{ or }&
  \mu_1 = 2\nu_2\,,\quad
  \mu_2 = \frac{2\nu_2^2}{\nu_1}\,,\quad
  \mu_4 = \frac{2\mu_3\nu_2}{\nu_1}\,,\quad
  \mu_5 = \frac{\mu_3\nu_2^2}{\nu_1^2}\,.
\label{FinSub:eq}
\end{eqnarray}
For the {\sc Lie} algebra spanned by these vector-fields we choose the
following basis ($j,k=1,\ldots,n$):
\begin{equation}
\begin{array}{c}
  L_{jk} = x_j\partial_{x_{\! k}} - x_{\! k}\partial_{x_j}\,,\quad
  H = \partial_t\,, \quad
  D = \sum_{j=1}^n x_{\! j} \partial_{x_{\! j}} + 2 t \partial_t
    -\sfrac{n}{2}  \partial_r
    +\sfrac{n\mu_1}{2\nu_1} \partial_s\,, \\[1mm]
  C = \sum_{j=1}^n {x_{\! j}}t\partial_{x_{\! j}} + t^2\partial_t
    - \sfrac{n}{2} t\partial_r
    -\left(\sfrac{1}{4\nu_1}\, \x^2-\sfrac{n\mu_1}{2\nu_1}\,t\right)
    \partial_s\,,\quad
  P_j = \partial_{x_{\! j}}\,,\\[1mm]
  B_j = t\partial_{x_{\! j}} - \sfrac{1}{2\nu_1} {x_{\! j}}
    \partial_s \,, \quad
  E = -\sfrac{1}{2\nu_1}\partial_s \,,\quad
  R = \partial_r  \,,\quad
  A = - t \partial_t +\left(\sfrac{2\nu_2}{\nu_1}r +s\right) \partial_s\,.
\end{array}
\label{VecBas:eq}
\end{equation}
This leads to the following nontrivial commutators
($j,k,l,m=1,\ldots,n$):
\begin{equation}
\begin{array}{c}
\ds\big[D, H\big] = -2 H\,,\quad
  \big[H, C\big] = D\,,\quad
  \big[D, C\big] = 2 C\,,\quad
  \big[H, B_j\big] = P_j\,, \\[1mm]
\ds\big[D, P_j\big] = - P_j\,,\quad
  \big[D, B_j\big] = B_j\,,\quad
  \big[C, P_j\big] = - B_j\,,\quad
  \big[P_j, B_k\big] = \delta_{jk} E\,, \\[1mm]
\ds\big[A, H\big] = H \,,\quad
  \big[A, C \big] = -C \,,\quad
  \big[A, E \big] = -E \,,\quad
  \big[A, R  \big] = 4\nu_2 E \,,\quad
  \big[A, B_j\big] = -B_j\,, \\[1mm]
\ds\big[L_{jk},P_l\big] = \delta_{kl}P_j - \delta_{jl}P_k\,,\qquad
  \big[L_{jk},B_l\big] = \delta_{kl}B_j - \delta_{jl}B_k\,,\\[1mm]
\ds\big[L_{jk},L_{lm}\big] = \delta_{kl} L_{jm}+\delta_{jm}L_{kl}
    -\delta_{jl}L_{km} -\delta_{km}L_{jl}\,.
\end{array}
\end{equation}
Now if both conditions \Ref{GalSub:eq} and \Ref{FinSub:eq} are
fulfilled, i.e.\ for the two parameter subfamily
\begin{equation}
  \mu_1 = 2\nu_2\,,\quad
  \mu_2 = \frac{2\nu_2^2}{\nu_1}\,,\quad
  \mu_3 = -\nu_1\,,\quad
  \mu_4 = -2\nu_2\,,\quad
  \mu_5 = -\frac{\nu_2^2}{\nu_1}\,,
\end{equation}
we get an $\frac{(n+1)(n+2)}{2}+5$ dimensional {\sc Lie} symmetry
algebra of the DG--equation \Ref{nse:eq}.
$\{H,D,C\}$ generate the subalgebra $sl(2,\rz)$, $\{P_j\}$ and
$\{B_j\}$ two
$n$-dimensional commutative subalgebras, and $\{L_{jk}\}$ the subalgebra
$so(n)$. Together these
generators span the {\sc Schr\"o\-dinger} algebra $sch(n)$.
Note, however, that the representation of $sch(n)$ differs from the
one obtained for the linear SE \cite{Nieder1,Hagen1} by those terms
proportional to $\mu_1$ in $D$ and $C$; the reason will be explained
in the section~\ref{cla:sec}.
Furthermore, the generator $R +4\nu_2 E$ is easily
seen to be independent of the rest, and $\{A,E\}$ span a one dimensional
affine algebra that acts on the $sl(2,\rz)$- and the boost-subalgebra of
the {\sc Schr\"odinger} algebra $sch(n)$. Hence, the structure of this
{\sc Lie} symmetry can be written as
\begin{equation}
  sym_3(n) = \Big(\aff(1)\splus sch(n) \Big)\oplus t(1)\,.
\label{sym2:eq}
\end{equation}

If only \Ref{GalSub:eq} is fulfilled, $A$ is not a generator of a {\sc Lie}
symmetry of \Ref{nse:eq}, and $E$ becomes the centre of the extension
$sch_e(n)$
of the {\sc Schr\"odinger} algebra.
Thus, we obtain for the five parameter subfamily \Ref{GalSub:eq} the finite
dimensional symmetry algebra
\begin{equation}
  sym_1(n) = sch_e(n) \oplus t(1)\,.
\label{sym1:eq}
\end{equation}
It is exactly the finite dimensional part of the maximal
{\sc Lie} symmetry algebra of the linear SE including real homogeneity but
excluding the (infinite dimensional) part due to the additivity of the
linear SE.

If on the other hand condition \Ref{GalSub:eq} is not fulfilled, but
\Ref{FinSub:eq} is, $C$ and $B_j$ are no longer generators of a {\sc Lie}
symmetry, the {\sc Schr\"odinger} subalgebra of $sym_3(n)$ is reduced to
a direct sum of $\aff(1)$ and the {\sc Euclid}ean algebra $e(n)$, and
we obtain as the {\sc Lie} symmetry algebra for the three parameter
subfamily \Ref{FinSub:eq}:
\begin{equation}
  sym_2(n) = \Big(\aff(1) \splus \left(\aff(1)\splus e(n)\right)\Big)
    \oplus t(1) \,.
\label{sym3:eq}
\end{equation}
However, if we impose no conditions at all,
the {\sc Lie} symmetry algebra is restricted to the intersection of
$sym_1(n)$ and $sym_2(n)$:
\begin{equation}
  sym_0(n) =  \Big(\aff(1)\splus e(n)\Big) \oplus t(2)\,.
\label{sym0:eq}
\end{equation}

As mentioned previously the exponential solution of \Ref{dphis:eq} and
\Ref{dphis1:eq},
\begin{eqnarray}
  \phi(\x,t,r,s) &=& \phi^{\sss (1)}(\x,t,r) e^{\lambda s}\,,
\label{phi_exp:eq}\\[2mm]
  \sigma(\x,t,r,s) &=& -\left(\frac{\mu_3+2\nu_1}{\nu_1}\,
    \lambda^{-1}+\frac{2\nu_2}{\nu_1} \right) \phi^{\sss (1)}(\x,t,r)
    e^{\lambda s}\,,
\label{sigma_exp:eq}
\end{eqnarray}
also leads to a finite dimensional {\sc Lie} symmetry.
The determining equation \Ref{d14:eq} implies that the $r$--dependence is
also exponential,
\begin{equation}
  \phi^{\sss (1)}(\x,t,r) = \phi^{\sss (2)}(\x,t) e^{\eta r}\,,\quad
  \eta=\frac{\mu_1}{\nu_1}\lambda-\frac{\mu_3+2\nu_1}{\nu_1}\,.
\end{equation}
The remaining determining equations are consistent with this solution in
three particular cases. One of them leads to the subfamily
\Ref{InfSub:eq}, another to an infinite dimensional {\sc Lie}
symmetry; both will be treated in the next section.
The third case implies $\lambda =
2\frac{\mu_3+\nu_1}{\mu_1-2\nu_2}$ and $\phi^{\sss (2)} = const$,
and restricts the parameters to
\begin{equation}
\begin{array}{c}
  \mu_2 = \frac{\mu_3(\mu_1+2\nu_2)^2 (\mu_3+2\nu_1) +
    8\mu_1\nu_1^2\nu_2}{
    8\nu_1(\mu_3+\nu_1)^2}\,,\quad
  \mu_4 = \frac{\mu_3(\mu_1+2\nu_2)}{2\nu_1}\,,\quad
  \mu_5 = \frac{\mu_3}{2\nu_1}\mu_2\,, \\
  \mu_1-2\nu_2\neq 0 \neq \mu_3+\nu_1\,.
\end{array}
\label{ExpSub:eq}
\end{equation}
Obviously these conditions are not consistent with \Ref{GalSub:eq} and
\Ref{FinSub:eq}, so we get the generator
\begin{equation}
  F = e^{\eta r+\lambda s} \left( \partial_r
    -\kappa\partial_s\right)\,,\quad
  \kappa= \frac{\mu_3(\mu_1+2\nu_2)+2\nu_1\nu_2}{2\nu_1(\nu_1+\mu_3)}\,,
\label{Fgen:eq}
\end{equation}
only in addition to generators of the {\sc Lie} symmetry $sym_0(n)$
\Ref{sym0:eq}.
If we replace $D$ by $D' = \sum_{j=1}^n x_{\! j} \partial_{x_{\! j}}
+ 2 t \partial_t$ the additional nontrivial commutation relations are
\begin{equation}
  \Big[ F, E\Big] = \frac{\lambda}{2\nu_1}F\,,\quad
  \Big[ F, R\Big] = -\eta F\,.
\end{equation}
Thus, $\{F,E,R\}$ span a subalgebra and we obtain as the maximal {\sc
  Lie} algebra in this case:
\begin{equation}
  sym_4(n) = \Big(t(2)\splus t(1)\Big)\oplus\Big(\aff(1)\splus e(n)
    \Big)\,.
\label{sym4:eq}
\end{equation}
The one parameter groups
of transformations $\Phi^X_\epsilon$ of the wavefunction
corresponding to the vector-fields $L_{jk},\,P_j,\,B_j,\,H_j,\,E,$ and
$R$ are well known from the linear SE, so we only give them for the
new generators $A$ and $F$ and the slightly modified $D$ and $C$:
\begin{equation}
\begin{array}{rcl}
\ds  \left(\Phi^D_\epsilon\psi\right)(\x,t)
  &=& \exp\left(-\epsilon \frac{ n}{2}
  +i\epsilon\frac{ n\mu_1}{2\nu_1}\right) \psi(\x e^{-\epsilon},
    t e^{-2\epsilon})\,, \\[2mm]
\ds \left(\Phi^C_\epsilon\psi\right)(\x,t)
  &=& (1-\epsilon t)^{-\frac{n}{2}}
    \exp\left(i(\frac{\epsilon}{2\nu_1(1-\epsilon t)} \x^2+\frac{n\mu_1
    }{\nu_1}\ln|1-\epsilon t|)\right) \psi(\frac{\x}{1-\epsilon
    t},\frac{t}{1-\epsilon t}) \,,\\[2mm]
\ds \left(\Phi^A_\epsilon\psi\right)(\x,t)
  &=& \exp\left(i\left((e^\epsilon-1)(\frac{2\nu_2}{\nu_1}
      \ln|\psi(\x,e^\epsilon t)| +\arg\psi(\x,e^\epsilon
      t))\right)\right)
  \psi(\x,e^\epsilon t)  \,,\\[2mm]
\ds \left(\Phi^F_\epsilon\psi\right)(\x,t)
  &=&
\begin{array}[t]{l}
  \sqrt{2\epsilon +e^{-\lambda\arg\psi(\x,t)}|\psi(\x,t)|^{-\eta}}
    |\psi(\x,t)|^{\frac{\eta}{2}}\, \exp\Big(\frac{\lambda}{2}
    \arg\psi(\x,t)\\[1mm]
  -i\frac{\kappa}{2}\Big(\ln(2\epsilon
        +e^{-\lambda\arg\psi(\x,t)}|\psi(\x,t)|^{-\eta})
        +\lambda\arg\psi(\x,t) \\[1mm]
  +\eta \ln|\psi(\x,t)|\Big)\Big) \psi(\x,t)\,.
\end{array}
\end{array}
\label{ExpCD:eq}
\end{equation}
The transformations generated by $C$ are called {\em projective\/}
\cite{RidWin1} or {\em conformal\/} \cite{Hagen1} transformations or
{\em expansions\/} \cite{Nieder1}, whereas $D$ generates
a scaling of the independent variables $(\x,t)$ called {\em dilations}.

\subsection{Infinite symmetries}\label{inf:sec}
As indicated in section~\ref{gen:sec} there are two cases in which we
obtain
an additional infinite dimensional {\sc Lie} symmetry. Let us treat
the case
$\alpha_j=\beta_k=0$ in equations \Ref{dphis:eq}, \Ref{dphis1:eq} first,
i.e.\ we restrict our calculations to the three parameter subfamily
\Ref{InfSub:eq}. This simplifies the determining equations
\Ref{d6:eq}--\Ref{d21:eq} considerably. Equations \Ref{d14:eq},
\Ref{d15:eq}, \Ref{d17:eq}, \Ref{d20:eq}, and \Ref{d21:eq} imply that
\begin{equation}
  \sigma(\x,t,r,s) = -\frac{2\nu_2}{\nu_1}\phi(\x,t,r,s)
    + r\sigma^{\sss (1)}(\x,t) +s\sigma^{\sss (2)}(\x,t)
    + \sigma^{\sss (0)}(\x,t)\,.
\end{equation}
{}From \Ref{d6:eq}--\Ref{d11:eq} we infer furthermore that $\phi$ and
$\sigma$ do not depend on $(\x,t)$ and that
$\xi^{\sss (j)}$ and $\tau$ lead to generators of the
symmetry $sym_0(n)$ \Ref{sym0:eq}, which have been discussed in the
previous section. Then \Ref{d12:eq} and \Ref{d20:eq}
imply that $\sigma^{\sss (1)}\equiv \sigma^{\sss (2)} \equiv 0$, and
\Ref{d14:eq} yields an equation for $\phi^{\sss (1)}$,
\begin{equation}
  \nu_1 \phi^{\sss (1)}_r -\mu_1\phi^{\sss (1)}_s = 0\,.
\end{equation}
Integrating this equation and neglecting the generators of complex
homogeneity $R$ and $E$, which we have already
discussed in the previous section, we obtain the generators of an
infinite dimensional {\sc Lie} symmetry
\begin{equation}
  Y_f = f(\mu_1r+\nu_1s)\Big(\partial_r
  -\sfrac{2\nu_2}{\nu_1} \partial_s \Big)
\label{Yinf:eq}
\end{equation}
If $\mu_1=2\nu_2$, the algebra $a^\infty =\{Y_f|f\in C^\infty(\rz)\}$ is
a representation of the infinite dimensional commutative {\sc Lie}
algebra $C^\infty(\rz)$. Otherwise it is a representation of the
infinite
dimensional {\sc Lie} algebra $\Vect(\rz)$ of vector-fields
on $\rz$, where the commutator is defined by $[f_1,f_2](z)
:= f_1(z) f_2^\prime(z) -f_2(z) f_1^\prime(z)$:
\begin{equation}
  \Big[ Y_{f_1}, Y_{f_2}\Big] = (\mu_1-2\nu_2) Y_{[f_1,f_2]}\,.
\label{Ycom:eq}
\end{equation}
In the commutative case condition \Ref{InfSub:eq} is
compatible with condition \Ref{FinSub:eq} of the finite algebra
$sym_2(n)$.
The nontrivial commutation relations of $Y_f$ with the
generators of $sym_2(n)$ are
\begin{equation}
  \Big[Y_f , R\Big] = -\mu_1 Y_{f^\prime}\,,\qquad
   \Big[Y_f , E\Big] = \frac{1}{2} Y_{f^\prime}\,,\qquad
  \Big[ Y_f, A\Big] = Y_{zf^\prime}\,.
\end{equation}
Still, $R+4\nu_2 E$ commutes with the rest of the algebra and for the
two parameter subfamily
\begin{equation}
  \mu_1 = 2\nu_2\,, \quad
  \mu_2 = 2\frac{\nu_2^2}{\nu_1}\,, \quad
  \mu_3 = -2\nu_1\,, \quad
  \mu_4 = -4\nu_2\,, \quad
  \mu_5 = -2\frac{\nu_2^2}{\nu_1}
\label{InfaSub:eq}
\end{equation}
we obtain the maximal {\sc Lie} algebra
\begin{equation}
  sym_2^a(n) = sym_2(n)\splus  a^\infty = \Big(\aff(1) \splus
  \left(\left(\aff(1)\splus e_n(n)\right) \oplus a^\infty \right)\Big)
  \oplus t(1)\,.
\label{sym3a:eq}
\end{equation}
The one parameter group of transformations generated by $Y_f$ can be
calculated explicitly in the commutative case:
\begin{equation}
  \left(\Phi^D_\epsilon\psi\right)(\x,t)
  =\exp\left(\epsilon f\left(2\nu_2\ln|\psi(\x,t)|
    +\nu_1\arg\psi(\x,t)\right) \left(1-\sfrac{2\nu_2}{\nu_1}\right)
  \right) \psi(\x,t)
\end{equation}
For the non commutative case $A$ is no longer a generator of a {\sc
  Lie} symmetry, and we obtain the maximal {\sc Lie} symmetry
\begin{equation}
  sym_0^a(n) = sym_0(n)\splus a^\infty = \Big(\aff(1)\splus e(n)\Big)
  \oplus \Big(t(1)\splus a^\infty\Big) \oplus t(1)
\label{sym0a:eq}
\end{equation}
of the three parameter subfamily \Ref{InfSub:eq}. Note that the finite
dimensional
algebra $sym_4(n)$ is contained in $sym_0^a(n)$.

Another infinite dimensional {\sc Lie} symmetry algebra is obtained
from the
exponential solution \Ref{phi_exp:eq} and \Ref{sigma_exp:eq}. To simplify
calculations we can allow $\lambda$ to be complex valued, as
the determining equations are linear.
We mentioned in the previous section that only in
three cases the exponential ansatz is  consistent with the
determining equations \Ref{d6:eq}--\Ref{d21:eq}; one case led to a
finite symmetry, another leads to the infinite symmetry already
discussed
in this section. The remaining case requires a restriction to the three
parameter subfamily
\begin{equation}
  \mu_1 = 2\nu_2\,,\quad
  \mu_3 = -\nu_1\,,\quad
  \mu_4 = -2\nu_2\,,\quad
  \mu_5 = -\frac{1}{2}\mu_2\,,\qquad
  \mu_2 \neq 2\frac{\nu_2^2}{\nu_1}\,,
\label{EhrSub:eq}
\end{equation}
and the coefficient of the exponential function is fixed by
\begin{equation}
  \lambda^2 = \frac{\nu_1^2}{4\nu_2^2-2\nu_1\mu_2} \,.
\label{lambda:eq}
\end{equation}
So we have to distinguish two cases.

If $\mu_2 < 2\frac{\nu_2^2}{\nu_1}$, $\lambda = \pm |\lambda|$ is real
and we get the generators
\begin{equation}
\begin{array}{rcl}
  Z_{\Phi_\pm} &=&\ds e^{-r}\bigg(
    \Phi_+ (\x,t) e^{|\lambda|(\frac{2\nu_2}{\nu_1}r+s)}
      \left( |\lambda|\partial_r
        -(1+\sfrac{2\nu_2}{\nu_1}|\lambda|)\partial_s\right)\\&& \ds
    +\Phi_- (\x,t) e^{-|\lambda|(\frac{2\nu_2}{\nu_1}r+s)}
      \left( |\lambda|\partial_r
        +(1-\sfrac{2\nu_2}{\nu_1}|\lambda|)\partial_s\right)\bigg)
\end{array}
\label{Z1inf:eq}
\end{equation}
where $\Phi_+$ ($\Phi_-$) is a (smooth) solution of
the forward (backward) heat equation with diffusion coefficient
$\sqrt{4\nu_2^2-2\nu_1\mu_2}$, if $\nu_1>0$ (The role of $\Phi_+$ and
$\Phi_-$ is interchanged if $\nu_1<0$):
\begin{equation}
  \partial_t \Phi_\pm \pm \sqrt{4\nu_2^2-2\nu_1\mu_2}\, \Delta\Phi_\pm
  = 0\,.
\label{heat:eq}
\end{equation}
The algebra $b^\infty =\{Z_{\Phi_\pm}|\Phi_\pm$ solution of
\Ref{heat:eq}$\}$ is commutative, and the
subfamily \Ref{EhrSub:eq} possesses also the finite dimensional
symmetry $sym_1(n)$, whose
generators act on the generators $Z_{\Phi_\pm}$. Hence, we obtain the
full
{\sc Lie} symmetry algebra
\begin{equation}
  sym_1^{b} (n) = sym_1(n) \splus b^\infty\,.
\label{sym1b:eq}
\end{equation}
In order to calculate the one parameter groups of transformations
corresponding
to the generators \Ref{Z1inf:eq}, we have to solve the system of ordinary
differential equations for the initial conditions $r(0)=r_{\sss 0}$ and
$s(0)=s_{\sss 0}$:
\begin{eqnarray}
\label{dr:eq}
  r^\prime(\epsilon) &=& e^{-r}|\lambda| \left(
    \Phi_+(\x,t) e^{|\lambda|(\frac{2\nu_2}{\nu_1}r+s)}
   +\Phi_-(\x,t) e^{-|\lambda|(\frac{2\nu_2}{\nu_1}r+s)}\right)   \\
\label{ds:eq}
  s^\prime(\epsilon) &=& e^{-r}\left((1-\sfrac{2\nu_2}{\nu_1}
    |\lambda|)
     \Phi_-(\x,t) e^{-|\lambda|(\frac{2\nu_2}{\nu_1}r+s)}
     -(1+\sfrac{2\nu_2}{\nu_1} |\lambda|)
    \Phi_+(\x,t) e^{|\lambda|(\frac{2\nu_2}{\nu_1}r+s)}
    \right)
\end{eqnarray}
We immediately deduce that
\begin{equation}
  \frac{d^2}{d\epsilon^2} e^{2r(\epsilon)} =
  8\Phi_+(\x,t)\Phi_-(\x,t) \lambda^2\,,
\end{equation}
so
\begin{equation}
  r(\epsilon) = \half\ln\left( 4\Phi_+\Phi_- |\lambda|^2 \epsilon^2
  +2 e^{r_{\sss 0}} |\lambda| (\Phi_+e^{\lambda(\frac{2\nu_2}{\nu_1}
    r_{\sss 0}+
    s_{\sss 0})} +\Phi_-e^{-\lambda(\frac{2\nu_2}{\nu_1} r_{\sss 0}+
    s_{\sss 0})}) \, \epsilon
  + e^{2r_{\sss 0}}\right)\,.
\label{r_sol1:eq}
\end{equation}
Now $s(\epsilon)$ is determined by equation \Ref{dr:eq}, (The sign of the
square root is fixed by the initial condition $s(0)=s_{\sss 0}$.)
\begin{equation}
  s(\epsilon) = -\frac{2\nu_2}{\nu_1} r(\epsilon) +\frac{1}{|\lambda|}
  \ln\left( \pm\sqrt{\left( \frac{r^\prime(\epsilon)e^{r(\epsilon)}}{
          2|\lambda|\Phi_+}\right)^2 -\frac{\Phi_-}{\Phi_+}} -
    \frac{r^\prime(\epsilon)e^{r(\epsilon)}}{2\lambda\Phi_+} \right) \,.
\label{s_sol1:eq}
\end{equation}
Together equations \Ref{r_sol1:eq} and \Ref{s_sol1:eq} constitute the
one parameter group of transformations.
Since $\Phi_\pm$ are solutions of a linear equation we can
set $\epsilon=\half$ in these formulas without loss of generality
and for every solution $\psi_{\sss 0} =\exp(r_{\sss 0}+is_{\sss 0})$
of the
DG--equation equations \Ref{r_sol1:eq} and \Ref{s_sol1:eq} produce
a new solution.
In particular, if we start with the trivial solution $\psi_{\sss
  0}\equiv 0$
of \Ref{nse:eq}, i.e. $r_0\to -\infty$, we obtain a transformation
mapping pairs of positive solutions $(\Phi_+,\Phi_-)$ of the forward and
backward heat equation to a solution of equation \Ref{nse:eq},
\begin{equation}
  \psi(\x,t) = \left(\Phi_+(\x,t) \Phi_-(\x,t)\right)^{\half(1
    -i\frac{2\nu_2}{\nu_1})} \exp\left(\frac{i}{2|\lambda|}
    \ln\left(\frac{\Phi_-}{\Phi_+}\right)\right)
\end{equation}
It is in fact an extension of the transformation found by {\sc
G.~Auberson} and {\sc P.C.~Sabatier} \cite{AubSab1} for the special
case $\nu_2=0$.

If on the other hand $\mu_2 >2\frac{\nu_2^2}{\nu_1}$, then $\lambda =
\pm i \Lambda^{-1}$,
$\Lambda=\sqrt{\frac{2\nu_1\mu_2-4\nu_2^2}{\nu_1^2}}$,
is imaginary and we have to ensure that the coefficients of the
vector-field
are real valued. In this case we get the generator
\begin{equation}
\begin{array}{rcl}
  Z_\Psi &=&  \frac{2}{\Lambda} e^{-r}|\Psi(\x,t)|\bigg(
    \sin\left(\frac{2\nu_2r+\nu_1 s}{\nu_1 \Lambda}
      -\arg\Psi(\x,t)\right)\partial_r \\[2pt]&&
    +\left(\Lambda \cos\left(\frac{2\nu_2r+\nu_1 s}{\nu_1 \Lambda}
      -\arg\Psi(\x,t)\right)
    -\frac{2\nu_2}{\nu_1}\sin\left(\frac{2\nu_2r+\nu_1 s}{\nu_1 \Lambda}
      -\arg\Psi(\x,t)\right)\right)\partial_s\bigg)
\end{array}
\label{Z2inf:eq}
\end{equation}
where $\Psi$ is a solution of the free linear {\sc  Schr\"odinger} equation
\begin{equation}
  i\dt \Psi = \nu_1\Lambda\,\Delta\Psi \,.
\label{se:eq}
\end{equation}
Again the algebra $c^\infty = \{Z_\Psi|\Psi$ solution
of \Ref{se:eq}$\}$ is commutative, and the maximal {\sc Lie}
symmetry algebra is
\begin{equation}
  sym_1^c (n) = sym_1(n) \splus c^\infty\,.
\label{sym1c:eq}
\end{equation}

$Z_\Psi$ can be integrated in analogy to the integration of $Z_{\Phi_\pm}$
above and we find the following solution:
\begin{equation}
\begin{array}{rcl}
  r(\epsilon) &=&\ds \half\ln\left( 4|\Psi|^2
    \frac{\epsilon^2}{\Lambda^2}
    +4 e^{r_{\sss 0}}|\Psi|\sin\Big(\Lambda^{-1}(\sfrac{2\nu_2}{\nu_1}
    r_{\sss 0}+
    s_{\sss 0})-\arg\Psi\Big)\,\frac{\epsilon}{\Lambda} + e^{2r_{\sss
        0}}\right)\,,
\\
  s(\epsilon) &=&\ds  \begin{array}[t]{l}\ds
\Lambda \arg\Psi -\frac{\nu_2}{\nu_1}\ln\left(
    4|\Psi|^2 \frac{\epsilon^2}{\Lambda^2}+4
    e^{r_{\sss 0}}|\Psi|\sin\Big(\Lambda^{-1}(\sfrac{2\nu_2}{\nu_1}
    r_{\sss 0}+
    s_{\sss 0})-\arg\Psi\Big)\,\frac{\epsilon}{\Lambda} + e^{2r_{\sss
        0}}\right) \\ \ds
    + \Lambda \arcsin\left( \frac{2|\Psi|\frac{\epsilon}{\Lambda} +
        e^{r_{\sss 0}} \sin\Big(\Lambda^{-1}(\sfrac{2\nu_2}{\nu_1}
        r_{\sss 0}+
    s_{\sss 0})-\arg\Psi\Big)}{\sqrt{ 4|\Psi|^2
      \frac{\epsilon^2}{\Lambda^2}
    +4 e^{r_{\sss 0}}|\Psi|
    \sin\Big(\Lambda^{-1}(\sfrac{2\nu_2}{\nu_1} r_{\sss 0}+
    s_{\sss 0})-\arg\Psi\Big)\,\frac{\epsilon}{\Lambda} + e^{2r_{\sss
        0}}}}\right) \,.
  \end{array}
\end{array}
\label{Z2_int:eq}
\end{equation}
Again we can set $\epsilon=\half$ without loss of generality
and for every solution $\psi_{\sss 0} =\exp(r_{\sss 0}+is_{\sss 0})$
of the
DG--equation \Ref{Z2_int:eq} produces a new solution. Here, the
trivial solution $\psi_{\sss 0}\equiv 0$ yields a linearizing
transformation
\begin{equation}
  N_{(\Lambda,\gamma)}(\psi) = \psi^{\half(1+\Lambda+i\gamma)}
    {\bar\psi}^{\half(1-\Lambda +i\gamma)} =
    |\psi|\, e^{i(\gamma\ln|\psi|+\Lambda\arg\psi)}\,,
\label{gt:eq}
\end{equation}
with $\gamma = -\frac{2\nu_2}{\nu_1}$.
We remark here that the definition \Ref{gt:eq} for non-integer
$\Lambda$ is only formal due to the ambiguity of the phase $\arg\psi$.
For integer values of $\Lambda$, however, it can be shown that the
transformation is continuous in the $L^2$-topology \cite{Luecke1}.

\section{Gauge Classification}\label{cla:sec}
In this section we first review some of the properties of the
set of transformations ${\cal N} = \{N_{(\Lambda,\gamma)} |
(\Lambda,\gamma)\in \dot\rz\times\rz\}$
that have been discussed in \cite{DoGoNa1}. As the transformations
\Ref{gt:eq}
leave the probability density $\rho$ invariant,
they were called {\em nonlinear gauge transformations\/}
We will then apply the results to a {\em gauge\/}
classification of the {\sc Lie} symmetries of the DG--equation
\Ref{nse:eq}.
\begin{enumerate}
\item
The set of these transformations is a realization of the
affine group $\Aff(1)$ in one dimension
\begin{equation}
  N_{(\Lambda_1,\gamma_1)}\circ N_{(\Lambda_2,\gamma_2)} =
  N_{(\Lambda_1\Lambda_2,\Lambda_1\gamma_2+\gamma_1)}\,,
\end{equation}
the inverse is therefore $N_{(\Lambda,\gamma)}^{-1} =
N_{(\Lambda^{-1},
  -\Lambda^{-1} \gamma)}$.
\item
The 8 parameter family \Ref{nse:eq} is invariant under this action
of the affine group $\Aff(1)$, and the change of parameters is given
by
\begin{equation}
\begin{array}{c}
\ds
\nu_1^\prime = \frac{\nu_1}{\Lambda}\,,\quad
\nu_2^\prime = -\frac{\gamma}{2\Lambda}\nu_1 +\nu_2\,,\\[3mm] \ds
\mu_1^{\,\prime} = -\frac{\gamma}{\Lambda}\nu_1 + \mu_1\,,\quad
\mu_2^{\,\prime} = \frac{\gamma^2}{2\Lambda}\nu_1-\gamma \nu_2
- \frac{\gamma}{2}\mu_1+\Lambda \mu_2\,,\quad \mu_3^{\,\prime}
 = \frac{\mu_3}{\Lambda}\\[3mm] \ds
\mu_4^{\,\prime}= -\frac{\gamma}{\Lambda}\mu_3 + \mu_4\,,\quad
\mu_5^{\,\prime} = \frac{\gamma^2}{4\Lambda}\mu_3
- \frac{\gamma}{2}\mu_4
+ \Lambda\mu_5\,,\quad
\mu_0^{\,\prime} = \Lambda \mu_0\,.
\end{array}
\label{pt:eq}
\end{equation}
\item
The concept of a gauge classification of the DG--equation can be taken
further. Since we have an action of a two dimensional group on
an eight dimensional space of parameters, an appropriate description
of the family \Ref{GDGE:eq} is by six parameters invariant under the
group action, and two gauge
parameters. A possible choice of gauge invariants is:
\begin{equation}
\begin{array}{c}
\iota_1 = \nu_1\mu_2 -\nu_2\mu_1\,,\quad
\iota_2 = \mu_1-2\nu_2\,,\quad
\iota_3 = 1 + \mu_3/\nu_1\,,\quad
\iota_4 = \mu_4-\mu_1\mu_3/\nu_1\,,\\[3mm]\ds
\iota_5 = \nu_1(\mu_2+2\mu_5)-\nu_2(\mu_1+2\mu_4)
    +2\nu_2^2\mu_3/\nu_1\,,\quad
\iota_0 = \nu_1\mu_0\,.
\end{array}
\label{gi:eq}
\end{equation}
Thus, for the free equation the five parameters
$\iota_1,\ldots,\iota_5$
determine an equivalence class of nonlinear Schr\"odinger equations.
\end{enumerate}

First we remark that the gauge transformations explain the
modifications in the
representation of the {\sc Schr\"odinger} group found in
section~\ref{fin:sec}: The transformation of the parameters
\Ref{pt:eq} shows that
there is always at least one nonlinear gauge transformation such that
$\mu_1 =0$.
So the modified projective transformations and dilations \Ref{ExpCD:eq}
for $\mu_1\neq 0$ are related to the usual ones ($\mu_1=0$) by a
nonlinear gauge transformation $N_{(1,\mu_1/ \nu_1)}$.

Furthermore, the conditions for the various maximal {\sc Lie} symmetries
of the DG--equation have to be invariant under $\cal N$ and we can
characterize the corresponding subfamilies by gauge invariants $\iota_j$.
Summarizing the results of section~\ref{lie:sec} and the relations
between
the various subfamilies we obtain the following picture:

\bigskip
\epsfbox{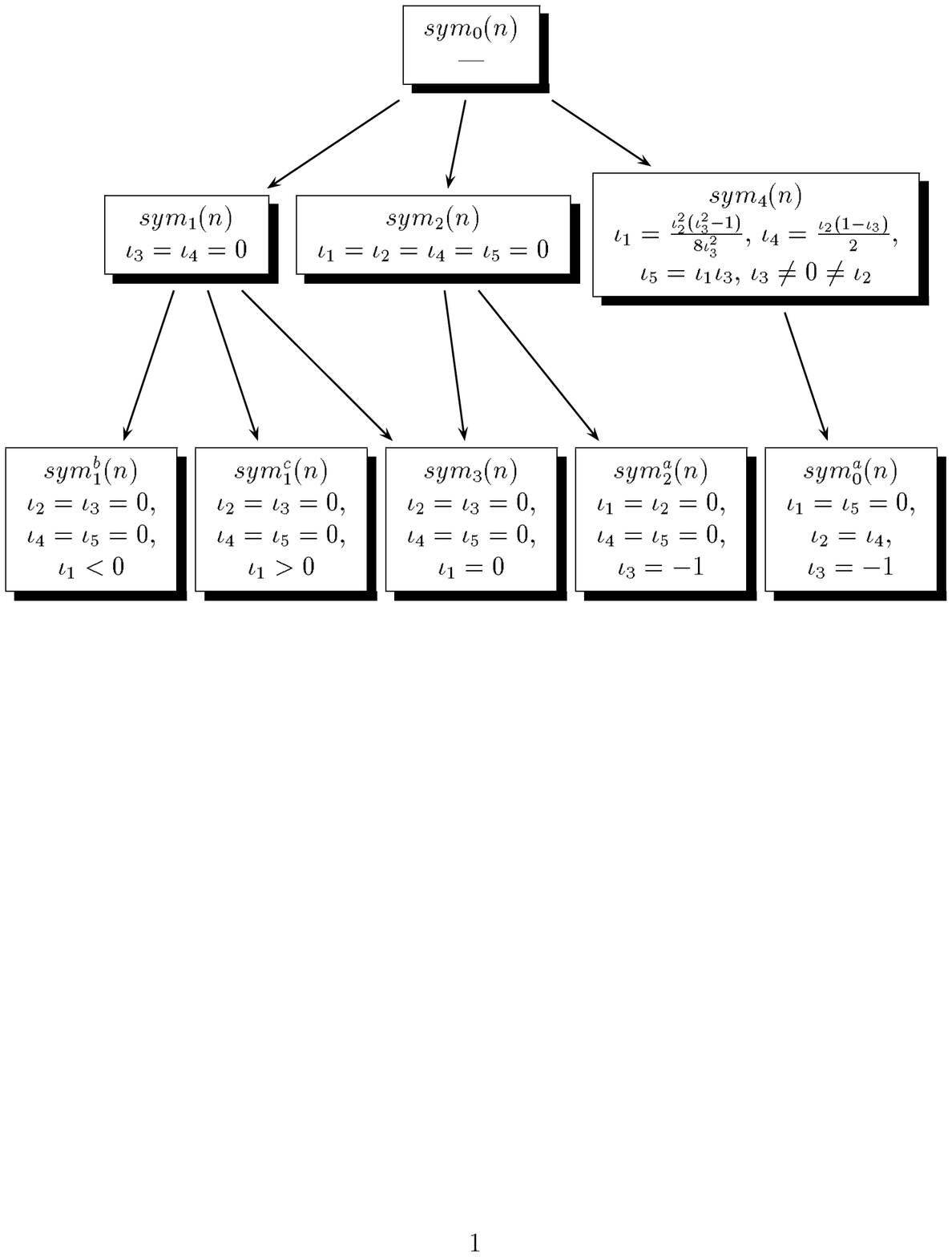}

\nopagebreak
\begin{description}
\item[{\bf Fig. 1}:] {\sc Lie} symmetries of the
  DG--equation. Subfamilies are
  characterized by gauge invariants and arrows indicate the
  subfamily-structure.
\end{description}
\section{Conclusions}
In this paper we have systematically discussed the {\sc Lie} symmetry
of the
$n+1$-dimensional DG--equation \Ref{GDGE:eq} in a convenient
parameterization \Ref{nse:eq}. Apart from the generators of the {\sc
  Schr\"odinger} group and complex homogeneity, we have found two new
generators $A$ and $F$ giving rise to the finite dimensional {\sc Lie}
symmetry algebras $sym_2(n)$, $sym_3(n)$ and $sym_4(n)$. Furthermore
we obtained four different types of infinite dimensional {\sc Lie}
symmetry algebras, $sym_0^a(n)$, $sym_2^a(n)$, $sym_1^b(n)$, and
$sym_1^c(n)$. The last two correspond to a local linearization of
a subfamily of the DG--equation to a forward and backward heat equation,
and a linear {\sc Schr\"odinger} equation, respectively. Due to the
systematic approach there are no other local linearizations of
the DG--equation.

One type of linearizing transformations leaves the whole family of
equations invariant, and was used to classify the subfamilies of the
various {\sc Lie} symmetry
algebras by gauge invariant parameters $\iota_j$ (Fig 1.).
The subfamily $\iota_2=\ldots=\iota_5=0$, $\iota_1\neq0$, is outstanding
because of its linearizability. The limiting case $\iota_1=0$ can be
transformed
into {\sc Euler}s equation \cite{AubSab1} and the methods to solve this
subfamily can be adapted to the subfamily
$\iota_1=\iota_2=\iota_4=\iota_5=0$
\cite{Renat}. The next candidate for a solvable subfamily seems to be the
subfamily $\iota_1=\iota_5=0$, $\iota_2=\iota_4$, $\iota_3=-1$, where the
infinite dimensional {\sc Lie} symmetry algebra might be used to
obtain a general solution. We postpone this for a future paper.

The results for the {\sc Schr\"odinger} invariant subfamily
\Ref{GalSub:eq}
are consistent (up to a nonlinear gauge transformation) with the results
of {\sc G.~Rideau} and {\sc P.~Winternitz} \cite{RidWin1} in $1+1$
dimensions, who classified all evolution
equations with $sch_e(1)$ {\sc Lie} symmetry algebra up to second
order in spatial
derivatives. The family \Ref{GalSub:eq} in the {\em gauge} $\nu_1=-1$ and
$\mu_1=0$ is contained in their equation (3.6)
for the special choice $S(u,v) = (-i\nu_2+\mu_2+\mu_5-\sfrac{1}{4}) u^2
+(i\nu_2-\mu_2+\half) v$.

Finally note that for the subfamily \Ref{EhrSub:eq} there
exists a {\sc Hamilton} functional generating the evolution equation
\cite{Natter1,DoGoNa1} and the corresponding DG--equation fits into
the scheme of a nonlinear quantum theory proposed by {\sc S.~Weinberg}
\cite{Weinbe2}. Furthermore a {\em consistent\/} concept of \lq nonlinear\rq\
observables (different from the one proposed by {\sc S.~Weinberg})
can be developed rigorously for the particular subfamily
$\mu_2 = \half\nu_1+2\frac{\nu_2^2}{\nu_1}$ of \Ref{EhrSub:eq}
establishing full equivalence between this model and linear quantum
mechanics \cite{Luecke1}.

\subsection*{Acknowledgments}
I would like to thank {\sc H.-D.~Doebner}, {\sc W.~Scherer}, and
{\sc R.~Zhdanov} for useful discussions. I am also grateful to the
organizers of the XXVIIth Symposium on Mathematical Physics in Toru\'n
for their hospitality.

\end{document}